\documentclass[sn-mathphys,Numbered]{sn-jnl}

\usepackage{graphicx}%
\usepackage{multirow}%
\usepackage{amsmath,amssymb,amsfonts}%
\usepackage{amsthm}%
\usepackage{geometry}
\usepackage{mathrsfs}%
\usepackage[title]{appendix}%
\usepackage{xcolor}%
\usepackage{textcomp}%
\usepackage{manyfoot}%
\usepackage{booktabs}%
\usepackage{algorithm}%
\usepackage{algorithmicx}%
\usepackage{algpseudocode}%
\usepackage{listings}%
\usepackage{breqn}
\usepackage{csquotes}
\usepackage{amssymb}

\theoremstyle{thmstyleone}%
\theoremstyle{thmstyletwo}%

\theoremstyle{thmstylethree}%

\begin{document}
\newcommand{\bu}{\mathbf{u}}
\title[Modal-based generalised quasilinear approximations for turbulent plane Couette flow]{Modal-based generalised quasilinear approximations for turbulent plane Couette flow}


\author*[1]{\fnm{Igor} \sur{A. Maia}}\email{igoriam@ita.br}

\author[1]{\fnm{Andr\'e} \sur{V. G. Cavalieri}}\email{andre@ita.br}
\equalcont{These authors contributed equally to this work.}

\affil*[1]{\orgdiv{Divis\~ao de Engenharia Aeroespacial}, \orgname{Instituto Tecnol\'ogico de Aeron\'autica}, \orgaddress{\street{Pra\c{c}a Mal. Eduardo Gomes, 50, Vila das Ac\'acias}, \city{S\~ao Jos\'e dos Campos}, \postcode{12228-900}, \state{S\~ao Paulo}, \country{Brazil}}}


\abstract{We study generalised quasilinear (GQL) approximations applied to turbulent plane Couette flow. The GQL framework is explored in conjunction with a Galerkin reduced-order model (ROM) recently developed by Cavalieri \& Nogueira (\textit{Phys. Rev. Fluids} 7, L102601, 2022), which considers controllability modes of the linearised Navier-Stokes system as basis functions, representing coherent structures in the flow. The velocity field is decomposed into two groups: one composed by high-controllability modes and the other by low-controllability modes. The former group is solved with the full nonlinear equations, whereas the equations for the latter are linearised. We also consider a new GQL framework wherein the linearised equations for the low-controllability modes are driven by nonlinear interactions of modes in the first group, which are characterised by large-scale coherent structures. It is shown that GQL-ROMs successfully recover the statistics of the full model with relatively high controllability thresholds and sparser nonlinear operators. Driven GQL-ROMs were found to converge more rapidly than standard GQL approximations, providing accurate description of the statistics with a larger number of linearised modes. This indicates that the forcing of linearised flow structures by large-scale coherent structures is an important feature of turbulence dynamics that should be considered in GQL models. The results presented here reveal that further model reductions are attainable with GQL-ROMs, which can be valuable to extend these models to larger Reynolds numbers.}




\maketitle

\section{Introduction}
\label{sec:intro}
Research on turbulent wall-bounded flows has always been permeated by attempts to find simplified descriptions of flow dynamics that would allow a better understanding of turbulence mechanisms and the design of flow control strategies. Much of the effort in that direction is concerned with the characterisation of a minimal set of coherent flow structures necessary for turbulence to sustain itself over long periods of time. This can be done, for instance, by truncating the flow domain to minimal flow units \cite{jimenez1991minimal}. The domain dimensions are chosen to be small enough so that the flow is periodic in the streamwise and spanwise directions. This removes some of the chaotic
aspect of the flow in space, while maintaining it in time. Typically, the minimal flow unit comprises a pair of high- and low-speed streaks of streamwise velocity, and a pair of counter-rotating streamwise vortices forming intermittently in time. The mechanisms of near-wall turbulence regeneration are now known to be intimately connected to the dynamics of the so-called \enquote{self-sustaining} processes involving those structures \cite{hamilton_kim_waleffe_1995}. 

Linear analysis then appears as a natural modelling framework, given the similarity of turbulent streaks and rolls with linear instability mechanisms found in transitional flows. In typical wall-bounded flows, the turbulent mean flow profile is linearly stable for a broad range of Reynolds numbers. But linear, transient amplification of disturbances can still occur due to the non-normality of the linearised Navier-Stokes operator, and linear tools are quite useful to understand some aspects of the flow, such as its stochastic response \cite{farrell1993a,farrell1993b}, optimal amplification mechanisms \cite{del2006linear, hwang2010linear, mckeon2010critical} and skin-friction generation \cite{lim2004singular,moarref2012model,blesbois2013pattern}. However, in a fully turbulent state, linear tools only offer a partial view of the flow dynamics; ultimately, one needs to add some degree of nonlinearity, however small, in order to obtain a more comprehensive view of turbulence. 

This prompted the development of \textit{quasilinear} (QL) approximations. The basic idea of this approach is to decompose the velocity field into two groups, following a given criterion. The equations for the first group are solved in their full nonlinear form, whereas in the second group self-interactions are either neglected or modelled. Numerous variants of quasilinear approximations have been used, differing in the criterion used for the decomposition or in the model for the self-interaction term. For instance, the effect of self-interactions has been modelled by an equivalent stochastic forcing term, as in the frameworks of stochastic structural stability theory (S3T) \cite{farrell2007structure,farrell2012dynamics} and direct statistical simulation (DSS) \cite{marston2008statistics,tobias2013direct}, or by a combination of stochastic forcing and an eddy-viscosity model \cite{hwang2020attached,skouloudis2021scaling}. Specifically in the case of wall-bounded turbulent flows, two QL variants have become noteworthy: the restricted nonlinear (RNL) approximation \cite{thomas2014self,farrell2017statistical} and generalised quasilinear (GQL) approximations \cite{marston2016generalized,tobias2017three,hernandez_yang_hwang_2022,hernandez_yang_hwang_2022_2}. In the RNL approach, the flow is decomposed into a zero-streamwise wavenumber component (streamwise mean), and Fourier modes with non-zero streamwise wavenumber. The equations for the latter are then linearised about the former, and self-interactions of non-zero wavenumbers are replaced by a stochastic forcing term. Although RNL models are capable of generating sustained turbulence and providing reasonable estimates of first- and second-order statistics \cite{thomas2014self}, they are severely limited in describing multi-scale energy exchanges and typical wavelength scalings \cite{hernandez_yang_hwang_2022}. This deficiency can be mitigated significantly in the GQL framework by allowing only a few modes with wavenumber other than zero to interact nonlinearly \cite{hernandez_yang_hwang_2022,hernandez_yang_hwang_2022_2}.

All of these approaches involve manipulating the equations of motion for the full Navier-Stokes system, usually in the framework of direct numerical simulations (DNS) or large-eddy simulations (LES). An alternative way to obtain simplified nonlinear models is to use Galerkin projections to derive reduced-order models (ROMs). In this approach, the governing (partial differential) equations are projected onto a basis formed by a reduced number of spatial modes, which are usually related to coherent flow structures. This results in a simplified system of ordinary differential equations that is solved for the temporal coefficients associated with the modes, which allows a subsequent reconstruction of the velocity field. Common examples of bases used in ROMs are Fourier modes and proper orthogonal decomposition (POD) modes. An early example of such a ROM is the 4-mode model of Waleffe \cite{waleffe1997self} for a wall-bounded flow forced by a streamwise body force, which allows a discretisation using Fourier modes. The model describes an interaction cycle between streaks and streamwise vortices, a key element of wall-bounded turbulent flows, but that was found not to lead to a chaotic regime. Other ROMs with larger bases were proposed by Eckhardt \& Mersmann \cite{EckhardtMersmann} (19 modes) and Moehlis \cite{Moehlis2004} (9 modes). Simulations of dynamic systems described by these models lead to a chaotic behaviour, but only over limited time horizons, typically much shorter than those found in DNS of minimal flow units. Later, Cavalieri \cite{CavalieriPRF2021} showed that typical turbulence lifetimes could achieve order-of-magnitude increases if the basis described interactions between streamwise vortices and streaks of different spanwise wavelengths. Based on that observation, Cavalieri \& Nogueira \cite{Cavalieri&Nogueira_PRF2022} recently developed ROMs for Couette flow whose modes, taken as controllability modes of the linearised Navier-Stokes system, represent flow structures of a few different streamwise/spanwise wavenumber pairs. The models, whose dimensions are of the order of a few hundred degrees of freedom, have long turbulence lifetimes and were found to match DNS statistics with reasonable accuracy.

In this work, we explore GQL approximations applied to a ROM of plane Couette flow. We consider the ROM of Cavalieri \& Nogueira \cite{Cavalieri&Nogueira_PRF2022}, and we assess whether it is possible to reduce the number of nonlinear interactions required to produce sufficiently accurate representations of turbulence dynamics. The ROM offers a simplified framework in which mode interactions can be discarded in a straightforward way via manipulation of the coefficients of nonlinear terms. Furthermore, instead of using a wavenumber threshold for linearisation, as done in previous studies \cite{marston2016generalized,  tobias2017three, hernandez_yang_hwang_2022, hernandez_yang_hwang_2022_2}, we consider a linearisation criterion based on the controllability of the modal basis, which can be determined prior to performing any simulation. The GQL approximation is thus applied in a modal sense, with coherent structures split into two groups according to their controllability. As will be seen, this criterion also acts as a filter that separates high-energy and low-energy modes. Linearisation is then applied to low-energy modes. A success in this strategy would lead to further model reduction, with a lower number of non-linear interactions in the ROM seen to sustain turbulence and reproduce flow statistics.

The remainder of the paper is organised as follows: in \S \ref{sec:model}, the GQL approximation is explained in detail, both for the full system and in the ROM. Results of GQL-ROMs with different linearisation criteria are presented in \S \ref{sec:results}. In \S \ref{sec:discussion} we discuss the main trends observed in the results, and, finally, some concluding remarks are provided in \S \ref{sec:conclusion}.

\section{Model}
\label{sec:model}

\subsection{Generalised quasilinear formulation}

We consider the generalised quasilinear framework developed by Marston \textit{et al}. \cite{marston2016generalized} and explored by Tobias \& Marston \cite{tobias2017three}, Hern\'andez \textit{et al.} \cite{hernandez_yang_hwang_2022} and Hern\'andez \textit{et al.} \cite{hernandez_yang_hwang_2022_2} to study the dynamics of wall-bounded turbulence. The approach starts from the incompressible Navier-Stokes equations,

\begin{equation}
\frac{\partial \bu}{\partial t} + \bu \cdot \nabla \bu = -\nabla p + \frac{1}{Re}\nabla^2\bu,
  \label{NS}
\end{equation}
with $\bu$ the velocity vector, $p$ the pressure and $Re$ the Reynold number. The flow geometry we consider is plane Couette flow with zero mean pressure gradient, and the Reynolds number is defined as $Re = U_wh/\nu$, where $U_w$ is the wall velocity, $h$ is the half-channel height and $\nu$ is the kinematic viscosity. In what follows, the velocity components are normalised by the wall velocity, and the streamwise, wall-normal and spanwise directions are denoted by $x$, $y$ and $z$, respectively. The GQL framework relies on the decomposition of the velocity field into two groups,

\begin{equation}
\bu = \bu_l + \bu_s,
\end{equation}
where, for the moment, subscripts $l$ and $s$ stand loosely for large and small. We then define two projection operators,

\begin{subequations}
\centering
\begin{eqnarray}
&\mathcal{F}_l[\bu] \equiv \bu_l,\\
&\mathcal{F}_s[\bu] \equiv \bu_s,
\end{eqnarray}
\end{subequations}
which are linear. Projecting the Navier-Stokes equations using $\mathcal{F}_l$ and $\mathcal{F}_s$ leads to,

\begin{equation}
\frac{\partial \bu_l}{\partial t} +\mathcal{F}_l \left[ \bu_l \cdot \nabla \bu_l\right] + \mathcal{F}_l \left[\bu_l \cdot \nabla \bu_s\right] + \mathcal{F}_l \left[\bu_s \cdot \nabla \bu_l\right] + \mathcal{F}_l \left[\bu_s \cdot \nabla \bu_s \right] = -\nabla p_l +\frac{1}{Re}\nabla^2\bu_l,
\end{equation}

\begin{equation}
\frac{\partial \bu_s}{\partial t} +\mathcal{F}_s \left[ \bu_l \cdot \nabla \bu_l\right] + \mathcal{F}_s \left[\bu_l \cdot \nabla \bu_s\right] + \mathcal{F}_s \left[\bu_s \cdot \nabla \bu_l\right] +\mathcal{F}_s \left[\bu_s \cdot \nabla \bu_s \right] = -\nabla p_s +\frac{1}{Re}\nabla^2\bu_s,
\end{equation}
which was obtained using the following property of the projection operators, $\mathcal{F}_l[\mathcal{F}_s[\cdot]] = 0, \ \mathcal{F}_s[\mathcal{F}_l[\cdot]] = 0$. $p_l$ and $p_h$ are defined so as to enforce $\nabla \cdot \bu_l = 0$ and $\nabla \cdot \bu_s = 0$, and $p = p_l +p_s$. It can be seen that the projection operators act as filters that lead to two systems of equations for the evolution of the $\bu_l$ and  $\bu_s$ components, which are coupled. The standard GQL approximation consists in discarding the nonlinear self-interaction terms $\mathcal{F}_s \left[\bu_s \cdot \nabla \bu_s \right]$ and $\mathcal{F}_s \left[\bu_l \cdot \nabla \bu_l \right]$, leading to the following system,

\begin{equation}
\frac{\partial \bu_l}{\partial t} +\mathcal{F}_l \left[ \bu_l \cdot \nabla \bu_l\right] + \mathcal{F}_l \left[\bu_s \cdot \nabla \bu_s \right] = -\nabla p_l +\frac{1}{Re}\nabla^2\bu_l,
\end{equation}

\begin{equation}
\frac{\partial \bu_s}{\partial t} + \mathcal{F}_s \left[\bu_l \cdot \nabla \bu_s\right] + \mathcal{F}_s \left[\bu_s \cdot \nabla \bu_l\right] = -\nabla p_s +\frac{1}{Re}\nabla^2\bu_s.
\end{equation}
Notice that the cross-interaction terms $\mathcal{F}_l \left[\bu_l \cdot \nabla \bu_s \right]$ and $\mathcal{F}_l \left[\bu_s \cdot \nabla \bu_l \right]$ have also been removed. This is a requirement for energy conservation in an incompressible flow. In Fourier space, the nonlinear term takes the form of interactions between modes with different wavenumbers that form triads. It follows from the energy balance equation that the global net energy transfer rate between wavenumbers, i.e the sum of nonlinear interactions across all wavenumbers, must be zero. However, it can also be shown that energy must be conserved not only in a global sense, but also within groups of triads formed by a given subset of wavenumbers \cite{schmid2012stability}. In physical space, this has the the following implication: the energy transferred from (into) the term $\mathcal{F}_s \left[\bu_l \cdot \nabla \bu_l \right]$ has to go to (come from) $\mathcal{F}_l \left[\bu_l \cdot \nabla \bu_s \right]$ and $\mathcal{F}_l \left[\bu_s \cdot \nabla \bu_l \right]$. In other words, defining $T$ as a net energy transfer into the $\bu_s$ or $\bu_l$ components, then $T(\mathcal{F}_s \left[\bu_l \cdot \nabla \bu_l \right]) + T(\mathcal{F}_l \left[\bu_l \cdot \nabla \bu_s \right])+ T(\mathcal{F}_l \left[\bu_s \cdot \nabla \bu_l \right])=0$. Therefore, if the term $\mathcal{F}_s \left[\bu_l \cdot \nabla \bu_l \right]$ is removed, terms $\mathcal{F}_l \left[\bu_l \cdot \nabla \bu_s \right]$ and $\mathcal{F}_l \left[\bu_s \cdot \nabla \bu_l \right]$ also need to be removed, so that the nonlinear term remains conservative. The final result of these simplifications is a system of linearised equations for $\bu_s$ about $\bu_l$, which is nonstationary. The equations for $\bu_l$, on the other hand, are affected by $\bu_s$ through the term $\mathcal{F}_l \left[\bu_s \cdot \nabla \bu_s \right]$, which can be understood as generalised Reynolds stresses.

\subsection{ROM}

We now describe how the GQL approximation is explored in the framework of ROMs. We consider the ROM developed by Cavalieri \& Nogueira \cite{Cavalieri&Nogueira_PRF2022} for plane Couette flow, in which velocity fluctuations about the laminar solution, $\bu'$, are subject to the following modal decomposition,

\begin{equation}
\bu'(x,y,z,t) = \sum_ja_j(t)\phi_j(x,y,z),
\label{modexp}
\end{equation}
where $a_j$ are temporal mode coefficients and the modes, $\phi_j(x,y,z)$, form an orthonormal basis that respects the continuity equation and the boundary conditions. Here we consider a basis derived from eigenfunctions of the controllability Gramian. These modes represent flow states that are most easily influenced by an external excitation in a linearised framework. Consider a generic linear, forced system,

\begin{equation}
\mathbf{\dot{x}} = \mathbf{A}\mathbf{x} + \mathbf{B}\xi(t),
\end{equation}
where $\mathbf{x}$ is the state, $\mathbf{A}$ is the linear operator and $\mathbf{B}$ is an operator describing the structure of the forcing term. $\xi$ is a zero-mean, Gaussian stochastic process, such that $\mathcal{E}\{\xi(t)\xi(t')\}=\mathbf{W}\delta(t-t')$, where $\mathcal{E}$ is the expectation operator, $\mathbf{W}$ is the forcing spatial covariance and $\delta$ is the Dirac delta function. For stochastically-forced systems, the dynamics of the state is best described by its covariance, $\mathbf{P}$. For such linear systems, under the assumption of a space-time white-noise forcing, i.e. $\mathbf{W}=\mathbf{I}$, the covariance becomes the controllability Gramian. Assuming that $\mathbf{A}$ is globally stable, the Gramian is given as \cite{farrell1993a,farrell1993b},

\begin{equation}
\mathbf{P}=\int_{0}^{\infty} e^{\mathbf{A}t}\mathbf{B}\mathbf{B}^{\dag}e^{\mathbf{A}^{\dag}t}\mathrm{d}t,
\end{equation}
where the superscript $^{\dag}$ denotes the adjoint. For a system in statistical steady-state, $\mathbf{P}$ can be computed from the following Lyapunov equation \cite{farrell1993a},

\begin{equation}
\mathbf{A}\mathbf{P} + \mathbf{P}\mathbf{A}^{\dag} + \mathbf{B}\mathbf{B}^{\dag}=0.
\end{equation}
The reader is referred to Jovanovic \& Bamieh \cite{jovanovic_bamieh_2005} for details of the operators $\mathbf{A}$, $\mathbf{A}^{\dag}$, $\mathbf{B}$ and $\mathbf{B}^{\dag}$ for the linearised Navier-Stokes system, doubly periodic in the streamwise and spanwise directions. The controllability modes for a given pair of streamwise and spanwise wavenumbers, $k_x$ and $k_z$, are then given by the eigendecomposition of $\mathbf{P}$,

\begin{equation}
\mathbf{P}\hat{\phi}_i = \sigma_i\hat{\phi}_i, \quad \sigma_1 \geqslant \cdots \geqslant \sigma_n \geqslant 0,
\end{equation}
and they are energy-ranked by their associated eigenvalues, $\sigma_i$, which are all real and positive. For a linear system forced with white noise, which is the case in the present derivation, these modes are equivalent to POD modes \cite{farrell1993a}. Therefore, one may question the choice of controllability modes as basis, instead of directly taking POD modes from data of the nonlinear system, as done in previous studies \cite{noack2003hierarchy, aubry1988dynamics, khoo2022sparse}. There are two main advantages in using controllability modes: first, they are derived from first principles, through the linearised operator. As such, computing them does not require any experimental or numerical flow data. Secondly, obtaining a large number of converged POD modes for a turbulent flow from experiments or simulations requires a significant amount of temporal data, which is necessary to converge two-point statistics. Therefore, for relatively large bases, higher POD modes can easily suffer from lack of convergence, which might produce a negative impact on the model. This issue does not concern controllability modes, as a large number of them can be computed at ease, provided a sufficient spatial discretisation is used. 

A normal-mode \textit{Ansatz} is associated with each controllability mode, such that,

\begin{equation}
\phi_i(x,y,z) = \hat{\phi_i}(y)e^{i(k_xx + k_zz)},
\end{equation}
where $k_x$ and $k_z$ are integer multiples of the fundamental wavenumbers, $\alpha = 2\pi/L_x$, $\beta = 2\pi/L_x$, respectively. We consider a ROM that includes combinations of streamwise wavenumbers $k_x/\alpha = 0,1,2$ and spanwise wavenumbers $k_z/\beta = -2,-1,0,1,2$. $n=24$ controllability eigenfunctions are taken for each combination of wavenumber pair. In order to work with real-valued modes, the real and imaginary parts of the modes are separated into two different modes, which differ only by a $\pi/2$ phase shift. This leads to 48 modes for each wavenumber pair, and a total number of modes of $N=600$. The modes are obtained with a discretisation of the linearised system using Fourier modes in streamwise and spanwise directions, and Chebyshev polynomials in the wall-normal direction, so as to provide spectral accuracy in the calculations of derivatives and integrals. The domain dimensions are $L_x = 2\pi h$ and $L_z = \pi h$, discretised with $N_x=14$, $N_y=65$, $N_z = 14$ points. The basis was computed at a lower Reynolds number, $Re=100$, than that used in this study. As discussed in \citet{Cavalieri&Nogueira_PRF2022}, bases computed at low $Re$ provide better quantitative agreements with DNS data. The lower $Re$ produces additional viscous effects, similar to an eddy-viscosity term, which improves the description of coherent structures in turbulent flows by linearised models, as also shown in previous studies \cite{morra2019relevance, pickering2021optimal}.

\begin{figure}
\centering
\includegraphics[trim=0cm 0cm 0cm 0cm, clip=true,width=\linewidth]{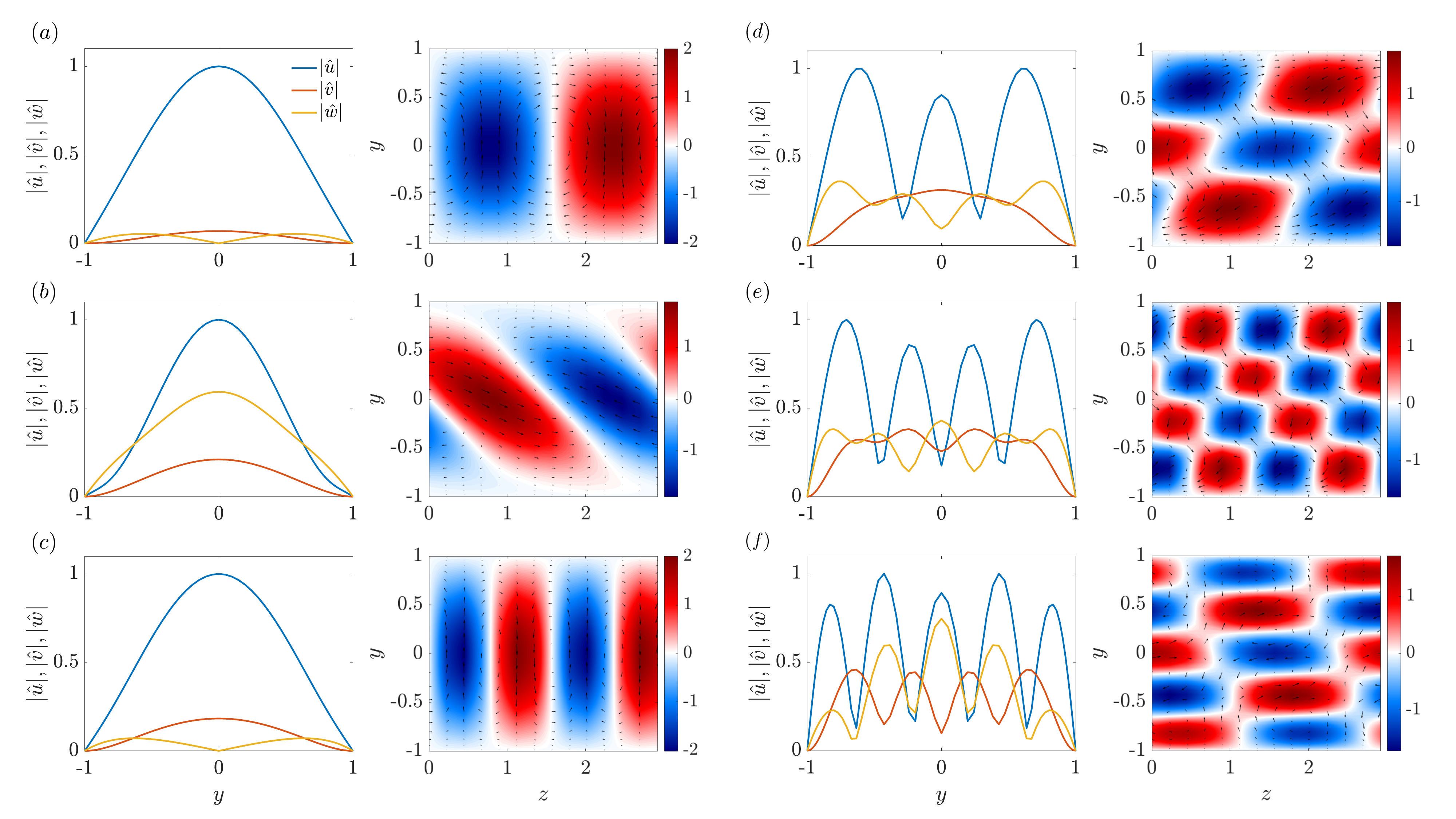}
\caption{Modal basis used in the ROMs, $\phi_j$. Panels (a) to (f) display modes 1, 3, 7, 50, 100 and 200, ranked in decreasing order of controllability (decreasing eigenvalues, $\sigma_j$). Left plots show the absolute values of the three velocity components, normalised by the maximum value of the streamwise component, $u$. Right plots show contours of the streamwise velocity component in a $y$-$z$ cut of the channel. Arrows represent the wall-normal, $v$, and spanwise, $w$, velocity components.}
\label{fig:1}
\end{figure}

The spatial structures of a few selected modes of the basis are displayed in figure \ref{fig:1}. The modes are ranked in descending order of controllability, given by their associated eigenvalues, $\sigma_j$. The most controllable modes, shown in panels (a)-(c), represent coherent structures associated with the lift-up effect. Notice how modes (a) and (c) display traits of the streak-roll mechanism: alternating patterns of low and high streamwise velocity, interspersed with streamwise vortices. The mode shown in (b) takes the shape of an oblique wave ($\alpha \neq 0$) with higher spanwise velocity components. They are associated with streamwise meandering motions, which are likely generated by streak instabilities \cite{hwang2016self}. Breakdown of these structures leads to the regeneration of the streamwise vortices and subsequent amplification of streaks, following the so-called self-sustaining processes \cite{hamilton_kim_waleffe_1995,waleffe1997self,jimenez_pinelli_1999,hwang_cossu_prl2010,flores2010hierarchy,hwang_2015,hwang_bengana_2016,de_giovanetti_sung_hwang_2017}.  Therefore, the most controllable modes are associated with key aspects of wall-bounded turbulence, and their nonlinear interactions are essential to its maintenance. Modes with lower controllability, illustrated in panels (d)-(f), are generally characterised by more oscillations in the wall-normal direction. They represent smaller-scale flow patterns, whose interpretability as turbulent coherent structures is less clear. As will be explained shortly, those are the modes concerned by the linearisation carried out in the GQL-ROM framework.

Inserting the modal expansion \ref{modexp} into the governing equations and taking the inner product with $\phi_i$, in a Galerkin procedure, leads to the following system of equations,

\begin{equation}
\frac{\mathrm{d}a_i}{\mathrm{d}t} = \frac{1}{Re}\sum_j L_{ij}a_j + \sum_j\tilde{L}_{ij}a_j + \sum_j\sum_k Q_{ijk}a_ja_k,
\label{rom}
\end{equation}
where, 

\begin{subequations}
\centering
\begin{eqnarray}
&L_{ij} = \left< \nabla^2\phi_j,\phi_i	\right>, \\
&\tilde{L}_{ij} =- \left< [(\phi_j \cdot \nabla)\bu_0 + (\bu_0 \cdot \nabla)\phi_j], \phi_i\right> \\
&Q_{ijk} = -\left< (\phi_j \cdot \nabla)\phi_k,\phi_i \right>,
\end{eqnarray}
\end{subequations}
are linear and nonlinear coefficients related to linear and non-linear terms of the Navier-Stokes system, $\mathbf{u}_0 = [y \ 0 \ 0 ]^T$ is the laminar flow solution and the symbol $\left< \cdot \right> $ denotes an inner product. This model has been validated by Cavalieri \& Nogueira \cite{Cavalieri&Nogueira_PRF2022} through comparisons with DNS results. Further validation of the dynamics was carried out by McCormack et al \cite{mccormack2023multi}, who verified that the ROM displays all features of scale interactions.

\subsection{Generalised quasilinear ROM}

We now consider a decomposition of the velocity fluctuation field, given in equation \ref{modexp}, into $\bu' = \bu_l'+\bu_s'$, where,

\begin{subequations}
\centering
\begin{eqnarray}
&\bu_l'(x,y,z,t) = \displaystyle \sum_{\sigma_j >\sigma_t} a_j^l(t)\phi_j^l(x,y,z), \\
&\bu_s'(x,y,z,t) = \displaystyle \sum_{\sigma_j \leqslant \sigma_t} a_j^s(t)\phi_j^s(x,y,z),
\end{eqnarray}
\end{subequations}
where $\sigma_t$ is a controllability threshold which is varied in order to produce GQL models with different number of linearised modes, as will be explained shortly. Defining projection operators analogous to those described in the previous section for the full Navier-Stokes equations, the reduced-order system can be split into two sets of ordinary differential equations,

\begin{equation}
\begin{split}
\frac{\mathrm{d}a_i^l}{\mathrm{d}t} = & {\frac{1}{Re}\sum_j L_{ij}a_j\rvert^{l} + \sum_j\tilde{L}_{ij}a_j\rvert^{l}} +\left (Q_{ijk}a_j a_k \rvert^{l,l,l}\right. \\
& +\left. Q_{ijk}a_j a_k \rvert^{l,l,s} + Q_{ijk}a_j a_k \rvert^{l,s,l} + Q_{ijk}a_j a_k \rvert^{l,s,s}\right),
\end{split}
\label{rom_ul}
\end{equation}

\begin{equation}
\begin{split}
\frac{\mathrm{d}a_i^s}{\mathrm{d}t} = & \frac{1}{Re}\sum_j L_{ij}a_j\rvert^{s} + \sum_j\tilde{L}_{ij}a_j\rvert^{s} + \sum_j\sum_k \left (Q_{ijk}a_j a_k \rvert^{s,l,l}\right. \\
& \left. Q_{ijk}a_j a_k \rvert^{s,l,s} + Q_{ijk}a_j a_k \rvert^{s,s,l} + Q_{ijk}a_j a_k \rvert^{s,s,s}\right),
\end{split}
\label{rom_us}
\end{equation}
where,

\begin{equation}
L_{ij}a_j\rvert^{l,s} =\left< \nabla^2\phi_j^{l,s},\phi_i^{l,s}\right>,
\end{equation}

\begin{equation}
\tilde{L}_{ij}\rvert^{l,s}  = \left< \left[(\phi_j^{l,s} \cdot \nabla)\bu_0 + (\bu_0 \cdot \nabla)\phi_j^{l,s}\right], \phi_i^{l,s}\right>.
\end{equation}
The triple superscripts in the quadratic terms denote the group to which belong the modes involved in a given triad in Fourier space. For instance, the interaction of two modes in the small set affecting a mode belonging to the large set is represented by, 

\begin{equation}
Q_{ijk}a_j a_k \rvert^{l,s,s} = -\left< (\phi_j^s \cdot \nabla)\phi_k^s,\phi_i^l \right>a_j^sa_k^s.
\end{equation}
The GQL approximation then amounts to discarding the $Q_{ijk}a_j a_k \rvert^{s,s,s}$ and $Q_{ijk}a_j a_k \rvert^{s,l,l}$ terms. Using the same arguments exposed above for energy conservation, removal of the term $Q_{ijk}a_j a_k \rvert^{s,l,l}$ (that represents interactions of two modes of the $\bu_l$ group transferring energy into a mode of the $\bu_s$ group), implies that $Q_{ijk}a_j a_k \rvert^{l,s,l}$ and $Q_{ijk}a_j a_k \rvert^{l,l,s}$ also have to be zeroed. In previous works \cite{marston2016generalized,hernandez_yang_hwang_2022,hernandez_yang_hwang_2022_2}, the velocity field was decomposed based on a wavenumber criterion, such that $u_l$/$u_s$ were composed of modes with wavenumber below/above a certain threshold. The dynamics of the full system are then approximated by successively increasing the cut-off wavenumber. Here we propose an alternative approach, wherein linearisation is based on a controllability criterion. Modes with high controllability, $\sigma_j >\sigma_t$, are grouped into $\bu_l$, whereas modes with eigenvalues equal or lower than this threshold are linearised. The rationale behind the controllability criterion has to do with the fact that the most controllable modes correspond to large coherent structures of the flow, as shown in figure 1. The nonlinear interactions of these structures underpin, to a great extent, the maintenance of turbulence through self-sustaining processes, as reviewed in the introduction. Low-controllability modes, on the other hand, have a less clear role in the nonlinear dynamics of turbulence, are expected to have low amplitude in the flow and are, therefore, more prone to linearisation. Assessing how high a controllability threshold can be set, or, in other words, how much of the basis can be linearised, without significant impacts on flow statistics is a central objective of the present work. This approach relies only on the modal basis, which is computed \textit{a priori}, without any input from the simulated full ROM. In this framework, linearisation impacts modes with different wavenumbers, the number of linearised modes for each wavenumber pair being, in general, different.

Furthermore, we also consider a new, modified GQL framework, in which $Q_{ijk}a_j a_k \rvert^{s,l,l}$ is kept in equation \ref{rom_us}. We refer to this approach as \enquote{driven GQL}, as opposed to the \enquote{standard} GQL explored in the aforementioned studies. This terminology comes from the fact that $Q_{ijk}a_j a_k \rvert^{s,l,l}$ acts as a source term that drives the equations for the evolution of $a_i^s$ (analogously to $\mathcal{F}_s \left[ \bu_l \cdot \nabla \bu_l \right]$ in the Navier-Stokes system). Hence, in the \enquote{driven} GQL approach, one has linear dynamics for $\bu_s$, but forced by nonlinear interactions involving $\bu_l$; the only neglected non-linear interactions are those involving exclusively $\bu_s$. For consistency, in this approach the terms $Q_{ijk}a_j a_k \rvert^{l,s,l}$ and $Q_{ijk}a_j a_k \rvert^{l,l,s}$ are also kept in equation \ref{rom_ul}. Figure \ref{fig:2} shows schematic diagrams of the different groups of triadic interactions in the full ROM. Triads in the bottom row, diagrams (e)-(h), are discarded in standard GQL, whereas only triads corresponding to group (h) are neglected in the driven GQL approach.

\begin{figure}
\centering
\includegraphics[trim=12cm 10cm 12cm 8cm, clip=true,width=\linewidth]{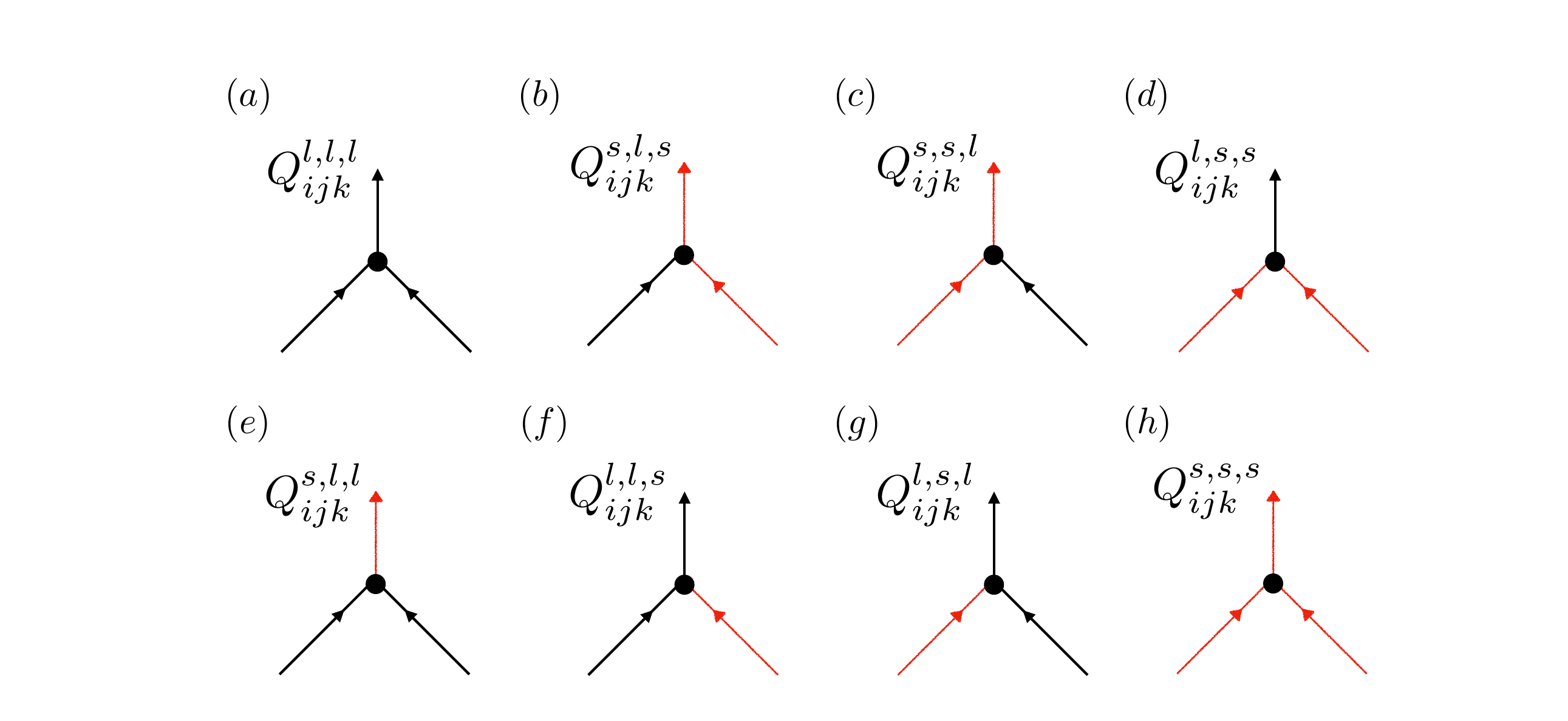}
\caption{Diagrams of the possible groups of triadic interactions involving the large- (black arrows) and small- (red arrow) controllability mode sets. In standard GQL, triads (e)-(h) are discarded; in driven GQL, only self-interactions of the small mode set, (h), are neglected.}
\label{fig:2}
\end{figure}

\section{Results}
\label{sec:results}

In what follows, we explore the performance of standard- and driven-GQL ROMs, subject to different linearisation criteria and thresholds, in providing simplified descriptions that retain nonetheless key aspects of the turbulence dynamics. The performance of GQL models will be systematically compared to that of the full ROM. The system of equations is initiated with a random condition and integrated in time using a standard 4th/5th Runge-Kutta scheme. Flow statistics are computed over 2000 time units, after excluding an initial transient. The Reynolds number of the simulations is $Re=1000$.

In figure \ref{fig:2}, first- and second-order statistics of the full ROM (i.e, without the GQL approximation), are compared to results from a DNS performed with the spectral code \textit{Dedalus} \cite{burns_etal_dedalus}, at the same Reynolds number and box dimensions. Table \ref{tab:dns_params} shows the flow and geometric parameters used in the DNS. Dealiasing is ensured in the DNS by multiplying the number of points in the streamwise and spanwise directions, $N_x$ and $N_z$, by a factor of 3/2. Statistics are expressed in wall units, $u^+ = u/u_\tau$, $v^+ = v/u_\tau$, $w^+ = w/u_\tau$, $y^+ = (y+1)u_\tau/\nu$. The mean flow profile is given as $U^+ = \left< u^+\right>_{x,z,t}$, where the symbol $\left< \cdot \right>_{x,z,t}$ denotes averaging in the streamwise and spanwise directions and in time. Throughout the remainder of the paper, the primes that denote velocity fluctuations are dropped for simplicity. The friction velocity is defined as $u_\tau = \sqrt{\tau_w/\rho}$, where $\tau_w$ is the shear stress at the wall and $\rho$ is the density. Both the mean flow and the Reynolds stresses are in reasonable agreement with the DNS, as previously reported by Cavalieri \& Nogueira \cite{Cavalieri&Nogueira_PRF2022}.

We quantify the accuracy of the different models used in the present study by an error metric, which is based on an $L^2$ norm of the difference in second-order statistics with respect to a reference case. For instance, the error in the description of the $\left< uu \right>^+_{x,z,t}$ profile is given by,

\begin{equation}
\epsilon_{\left<uu\right>^+} = \frac{|| \left< uu \right>_{ref}^+ - \left< uu \right>^+||_{2}}{||\left< uu \right>_{ref}^+||_{2}},
\label{err_def}
\end{equation}
where the averaging subscripts are omitted for simplicity and $\left< \cdot \right>_{ref}^+$ denotes the statistics of the reference case. Errors associated with the other components of the Reynolds stress tensor are computed in an analogous manner. Table \ref{tab:rom_params} displays the errors obtained with the full ROMs, considering the DNS as a reference case. Based on these values, we define an error threshold, $\overline{\epsilon}$, given as the average of the four errors, $\epsilon_{\left< uu \right>^+}$, $\epsilon_{\left< vv \right>^+}$, $\epsilon_{\left< uv \right>^+}$, $\epsilon_{\left< ww \right>^+}$, which will be used later to assess the accuracy of the GQL approximations. This threshold is of course arbitrary, but it reflects a reasonable level of accuracy one should expect from these models. It should be kept in mind that, given their (very) low dimensionality, perfect agreement with the data is not to be expected. Of course one can always aim at better accuracy, at the cost of increasing the size of the basis or keeping more nonlinear interactions. Furthermore, despite the (relatively small) errors, the full ROMs have been shown to correctly describe fundamental aspects of turbulence dynamics, including inter-scale interactions and self-sustaining processes. Therefore, it seems reasonable to adopt the same threshold as a good compromise between simplicity and accuracy for GQL-ROMs. The Reynolds number in the simulations is fixed at $Re=1000$, but the friction Reynolds number, $Re_\tau = u_\tau h/\nu$, fluctuates in time. As Couette flow is shear-driven, an accurate estimation of the mean friction Reynolds number reflects a correct description of the energy input to the system. It is thus an important parameter to assess the accuracy of the model. The average $Re_\tau$ computed from the full ROM is found to match that of the DNS to within 5\%.

\begin{figure}
\centering
\includegraphics[trim=4.5cm 9.5cm 10cm 8cm, clip=true,width=\linewidth]{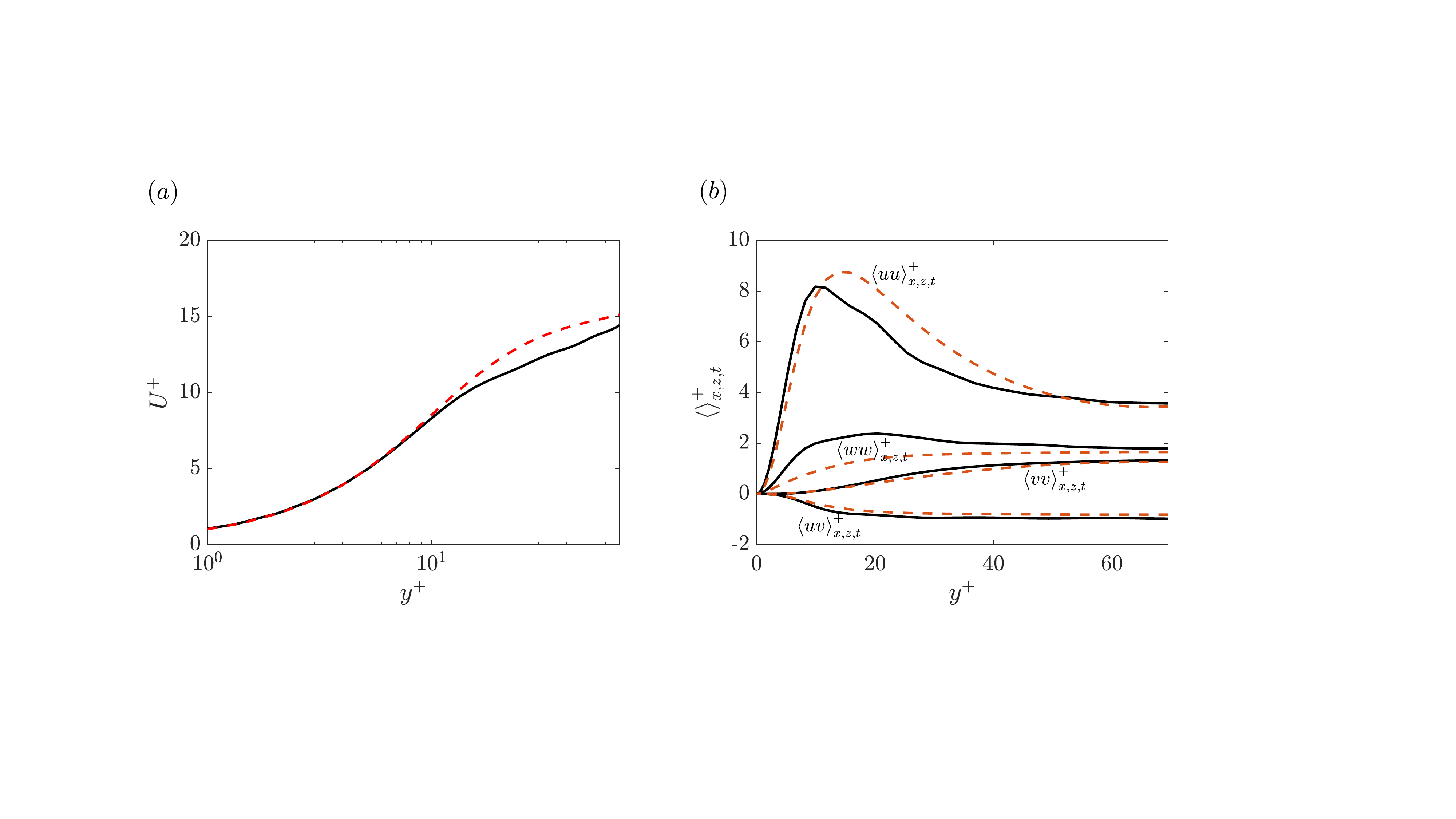}
\caption{Comparison of Couette flow statistics computed with DNS (red dashed lines) and the full ROM (black solid lines) at $Re=1000$. (a) Mean flow profiles; (b) second-order statistics.}
\label{fig:2}
\end{figure}

\begin{table}
\centering
  \caption{DNS parameters.}
    \label{tab:dns_params}
  \begin{tabular}{ c c c c c c c c c }
  \hline\noalign{\smallskip}
	$Re$ & $Re_\tau$ & $L_x$ & $L_z$  & $N_x$ & $N_y$ & $N_z$ & $\Delta x^+$ & $\Delta z^+$\\ 
	1000 & 66.2 & $2\pi$ & $\pi$ & $48$ & $68$ & $48$ & $8.67$ & $4.33$ \\
\noalign{\smallskip}\hline
  \end{tabular}
\end{table}

\begin{table}
\centering
  \caption{Parameters of the ROM simulations and errors in second order statistics with respect to the DNS. An error threshold, $\overline{\epsilon}$, is defined as the average of the errors in the Reynolds stresses, $\epsilon_{\left< uu \right>^+}$, $\epsilon_{\left< vv \right>^+}$, $\epsilon_{\left< uv \right>^+}$, $\epsilon_{\left< ww \right>^+}$.}
    \label{tab:rom_params}
\begin{tabular}{ c c c c c c c c} 
  \hline\noalign{\smallskip}
	$N$ & $Re$ &$Re_\tau$ & $\epsilon_{\left< uu \right>^+}$ & $\epsilon_{\left< vv \right>^+}$ & $\epsilon_{\left< uv \right>^+}$  & $\epsilon_{\left< ww \right>^+}$ & $\overline{\epsilon}$ \\
	\hline
	600 & 1000 & 69.4 & 0.02 & 0.05 & 0.08 & 0.26 & 0.1\\
\noalign{\smallskip}\hline
\end{tabular}
\end{table}

\subsection{Standard GQL}
\label{sec:std_gql}

\begin{figure}
\centering
\includegraphics[trim=12cm 12cm 11cm 8cm, clip=true,width=\linewidth]{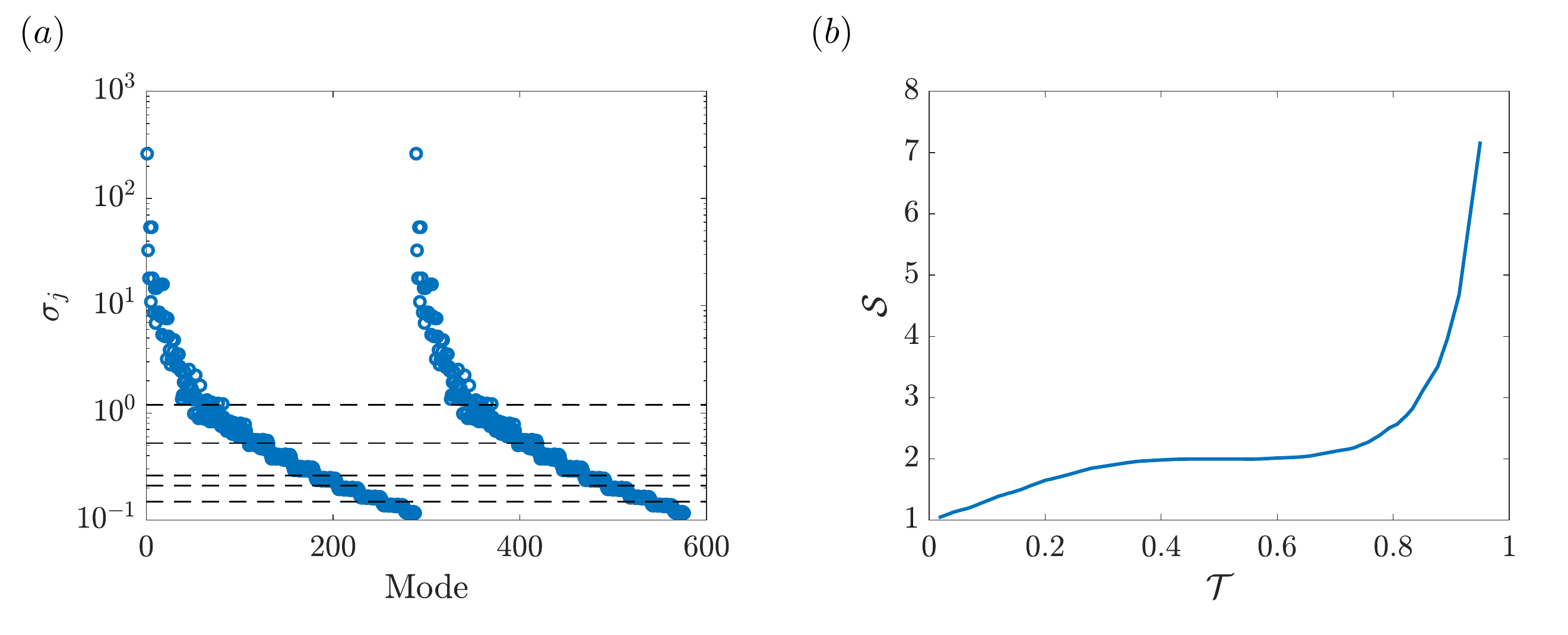}
\caption{Standard GQL with controllability criterion. (a) Eigenvalues of the controllability Gramian associated with the modes used in the current basis. Horizontal dashed lines represent five selected controllability thresholds, $\sigma_t$, with associated values of linearisation parameter $\mathcal{T} = 0.76, 0.56, 0.36, 0.28, 0.12$. Statistics associated with the selected thresholds are displayed presented in detail in figures \ref{fig:5} and \ref{fig:6}. (b) Sparsity index of the nonlinear term for different GQL approximations, as a function of linearisation threshold.}
\label{fig:4}
\end{figure}

\begin{figure}
\centering
\includegraphics[trim=12cm 12cm 11cm 8cm, clip=true,width=\linewidth]{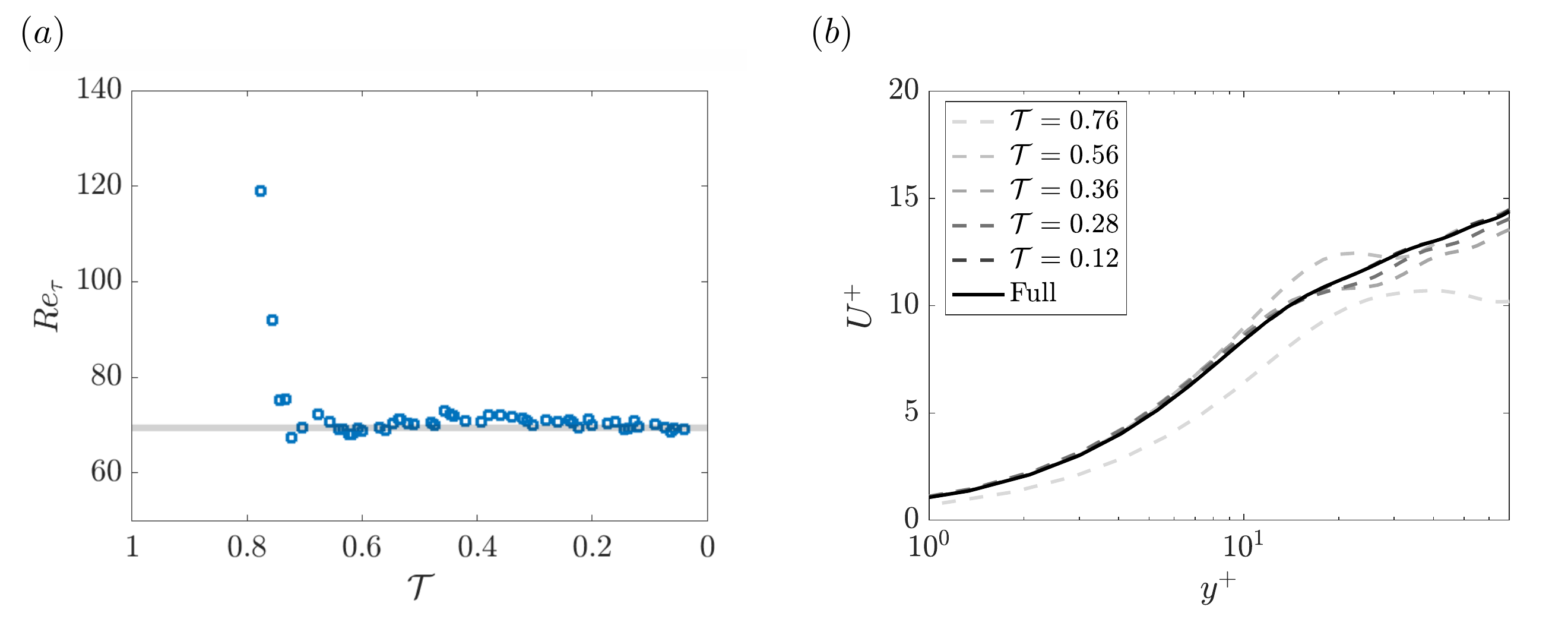}
\caption{Friction Reynolds number (a) and mean flow velocity profiles (B) for GQL models with different controllability threshold. The gray-shaded are in (a) represents the average reference Reynolds number, computed with 95\% confidence level from simulations of the full ROM. Results for five selected controllability thresholds are shown in (b).}
\label{fig:5}
\end{figure}

\begin{figure}
\centering
\includegraphics[trim=10cm 1cm 14cm 1cm, clip=true,width=\linewidth]{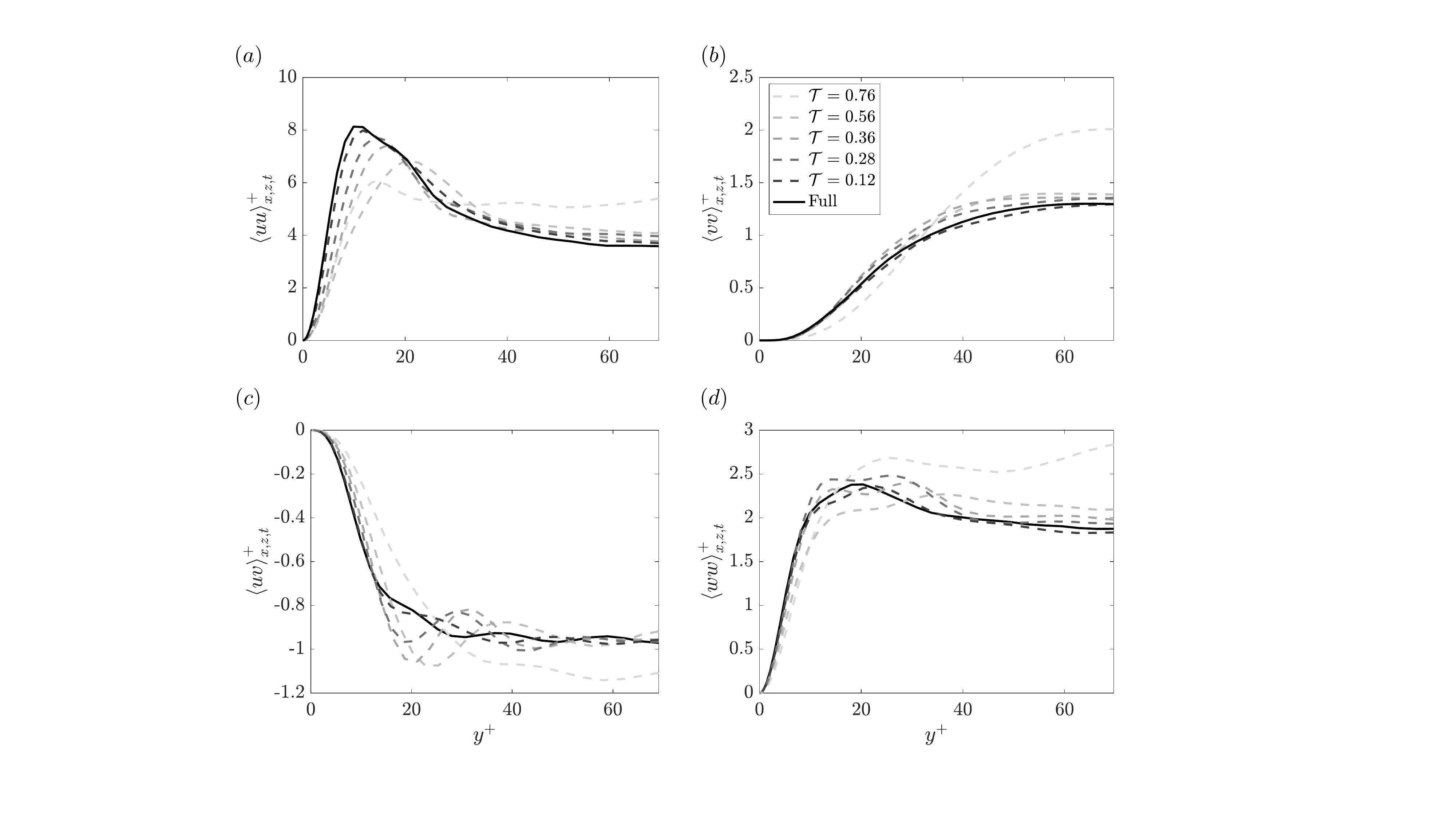}
\caption{Second-order statistics of different standard GQL-ROMs for selected controllability thresholds.}
\label{fig:6}
\end{figure}

We first consider standard GQL approximations. Figure \ref{fig:4} shows eigenvalues of the controllability Gramian, $\sigma_j$, associated with the modes used in the basis. The eigenvalues span roughly three orders of magnitude and are doubled. This is due to the fact that controllability eigenfunctions are split into two modes containing their real and imaginary parts, as explained above. We define a linearisation parameter, $\mathcal{T}$, associated with a given controllability threshold, $\sigma_t$. This parameter is given as the ratio between the number of linearised modes and the total number of modes in the basis, $\mathcal{T}=N_s/N$. It varies between 0 and 1; low values mean few modes are linearised, whereas values close to 1 indicate strong linearisation. Five selected thresholds, corresponding to $\mathcal{T} = 0.76, 0.56, 0.36, 0.28, 0.12$ are represented by dashed lines in figure \ref{fig:4}(a), whose associated statistics are displayed in figures \ref{fig:5} and \ref{fig:6}. The model reduction achieved with GQL approximations can be assessed via inspection of the nonlinear tensor, $Q_{ijk}$. In order to do that quantitatively, we define a sparsity parameter,

\begin{equation}
\mathcal{S} = \frac{N_{nz_{(Full)}}}{N_{nz_{(GQL)}}},
\end{equation}
where $N_{nz_{(GQL)}}$ and $N_{nz_{(Full)}}$ are the number of non-zero elements in the quadratic term in the GQL and full ROMs, respectively. Notice that the quadratic coefficients, $Q_{ijk}$, in the ROM are already sparse, as only modes satisfying a triadic rule have non-zero, nonlinear interactions; the sparsity parameter shows how much additional sparsity is obtained through the GQL approximation. By definition, $\mathcal{S}\geqslant 1$, and the sparsity parameter increases with increasing $\mathcal{T}$, as more modes are linearised. Ideally, from a modelling perspective, it is desirable to design a ROM with the highest possible $\mathcal{S}$, while keeping an accurate description of turbulence dynamics.

Simulations of GQL models were carried out for different levels of linearisation, starting from strongly linearised systems (high $\sigma_t$, high $\mathcal{T}$) and converging towards the full ROM ($\mathcal{T} \to 0$). The average friction Reynolds number of different GQL models is shown in figure \ref{fig:5}. The gray-shaded are represents the reference $Re_\tau$, computed with 95\% confidence level from 50 simulations of the full ROM. For high values of $\mathcal{T}$ (severe linearisation), $Re_\tau$ is significantly overestimated. As $\mathcal{T}$ decreases, $Re_\tau$ approaches that of the full ROM, although the convergence is not monotonic. In figure \ref{fig:5}(b), mean flow profiles of GQL ROMs for selected linearisation thresholds (highlighted in figure \ref{fig:4}(b)) are compared to the full ROM. It can be seen that the match with the mean profile of the full model improves substantially when $Re_\tau$ approaches the reference value, which occurs for $\mathcal{T} \gtrsim 0.6$.

Second-order turbulence statistics for the same thresholds are displayed in figure \ref{fig:6}. There is a clear trend of recovering the correct statistics with decreasing $\mathcal{T}$, with both the magnitude and the peak position converging to those of the full model. Similar observations were made by Hern\'andez \textit{et al.} \cite{hernandez_yang_hwang_2022} and Hern\'andez \textit{et al.} \cite{hernandez_yang_hwang_2022_2} using a wavenumber criterion to perform linearisation.

\begin{figure}
\centering
\includegraphics[trim=0cm 0cm 0cm 0cm, clip=true,width=0.5\linewidth]{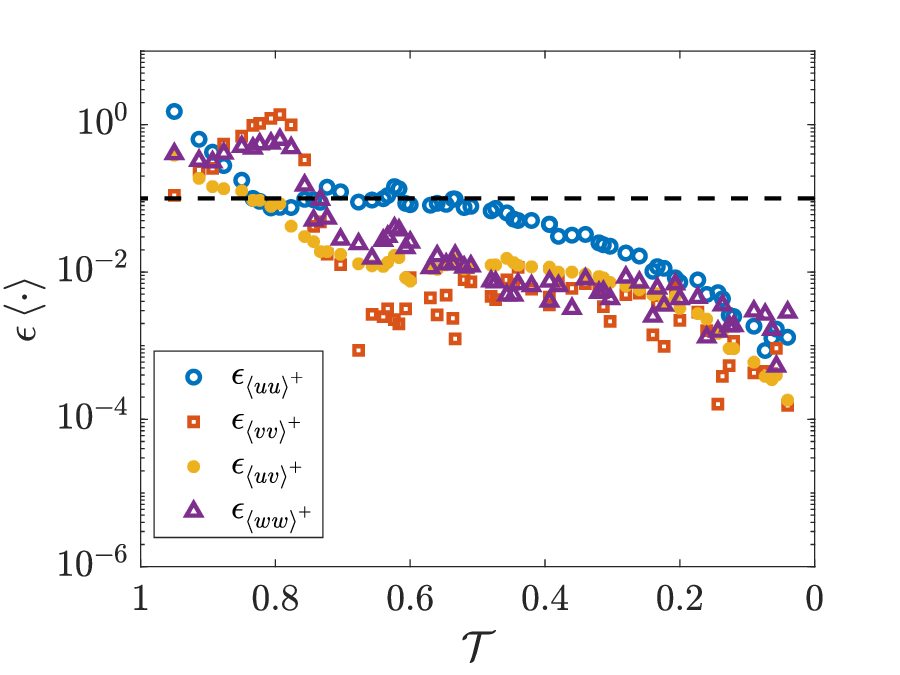}
\caption{Rms errors in the second-order statistics of the standard GQL models with respect to the full ROM.}
\label{fig:7}
\end{figure}

The errors in the characterisation of second-order statistics of the GQL-ROMs are quantified by the metric defined in equation \ref{err_def}, but this time with the reference values being that of the full ROM. Figure \ref{fig:7} shows the errors associated with the different components of the Reynolds tensor as a function of $\mathcal{T}$. The four errors are found to decrease in a quasi monotonic manner. The black-dashed line corresponds to the value of $\overline{\epsilon}$, the average error of the full ROM with respect to the DNS. We take this value as a criterion for accuracy: GQL-ROMs are consider accurate if all four errors with respect to the full ROM are below $\overline{\epsilon}$. This level of accuracy is reached for $\mathcal{T} \approx 0.6$, which is about the same threshold required to recover the correct friction Reynolds number, and the sparsity index associated with such a model is $\mathcal{S} = 2.01$. This implies that for the selected threshold half of the non-linear interactions of the original ROM are neglected without a significant loss of accuracy.

\subsection{Driven GQL}

\begin{figure}
\centering
\includegraphics[trim=12cm 12cm 11cm 8cm, clip=true,width=\linewidth]{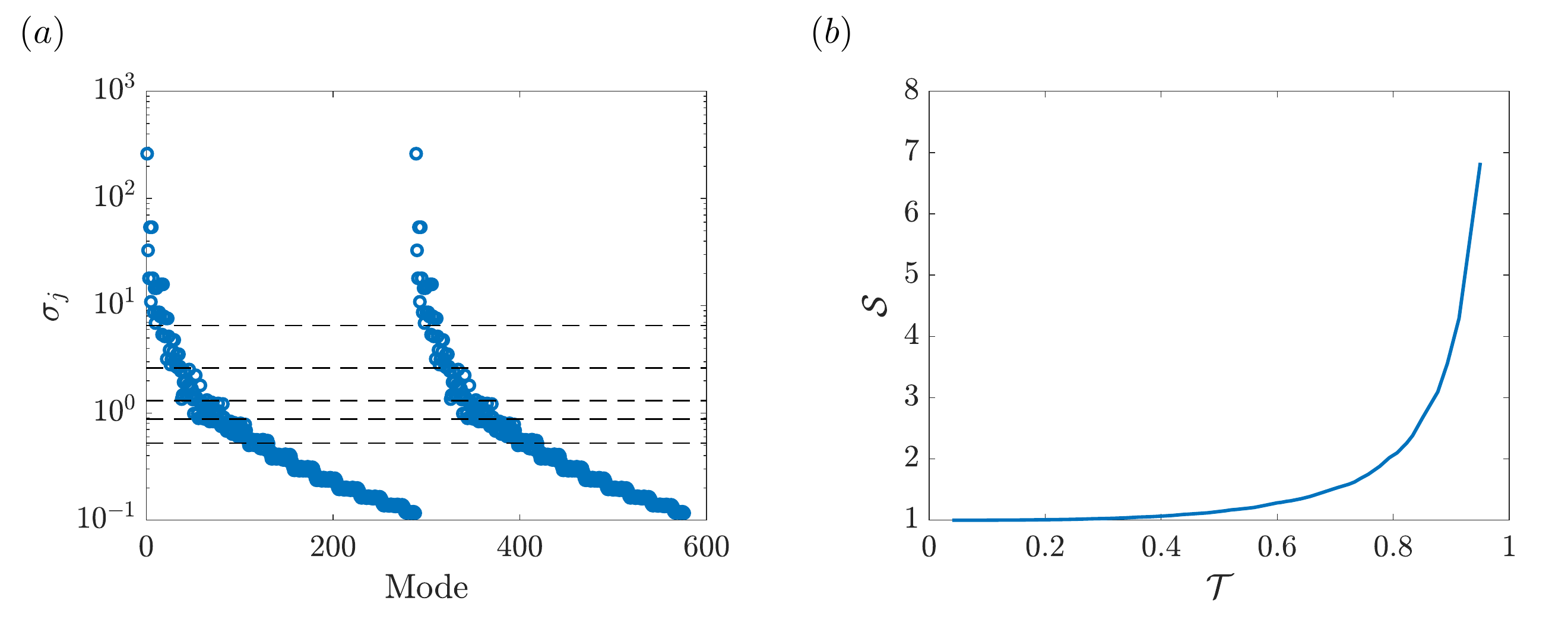}
\caption{Driven GQL with controllability criterion. (a) Eigenvalues of the controllability Gramian associated with the modes used in the current basis. Horizontal diagonal lines represent five linearisation thresholds, $\mathcal{T}= 0.91, 0.85, 0.78, 0.70, 0.56$, applied to ROMs whose statistics are displayed in figures \ref{fig:9} and \ref{fig:10}. (b) Sparsity index of the nonlinear term for different GQL approximations, as a function of controllability threshold.}
\label{fig:8}
\end{figure}

Next we consider the driven GQL approach, wherein terms $Q_{ijk}a_j a_k \rvert^{s,l,l}$, $Q_{ijk}a_j a_k \rvert^{l,l,s}$ and $Q_{ijk}a_j a_k \rvert^{l,s,l}$ are kept, and only $Q_{ijk}a_j a_k \rvert^{s,s,s}$ is discarded. In this approach, the term $Q_{ijk}a_j a_k \rvert^{s,l,l}$ drives the equations  for $a_i^s$. Figure \ref{fig:8} shows the sparsity index for different controllability thresholds. In this approach, as more terms are retained in the equations, the sparsity index decays faster with decreasing $\mathcal{T}$. The GQL models reach the same sparsity of the full ROM with $\mathcal{T} \approx 0.36$, whereas in the standard GQL approach this for values of $\mathcal{T}$ close to zero. Controllability eigenvalues are shown again for completeness in figure \ref{fig:8}(a), and five thresholds, different from those selected previously, are chosen to be studied in more detail.

\begin{figure}
\centering
\includegraphics[trim=12cm 12cm 11cm 8cm, clip=true,width=\linewidth]{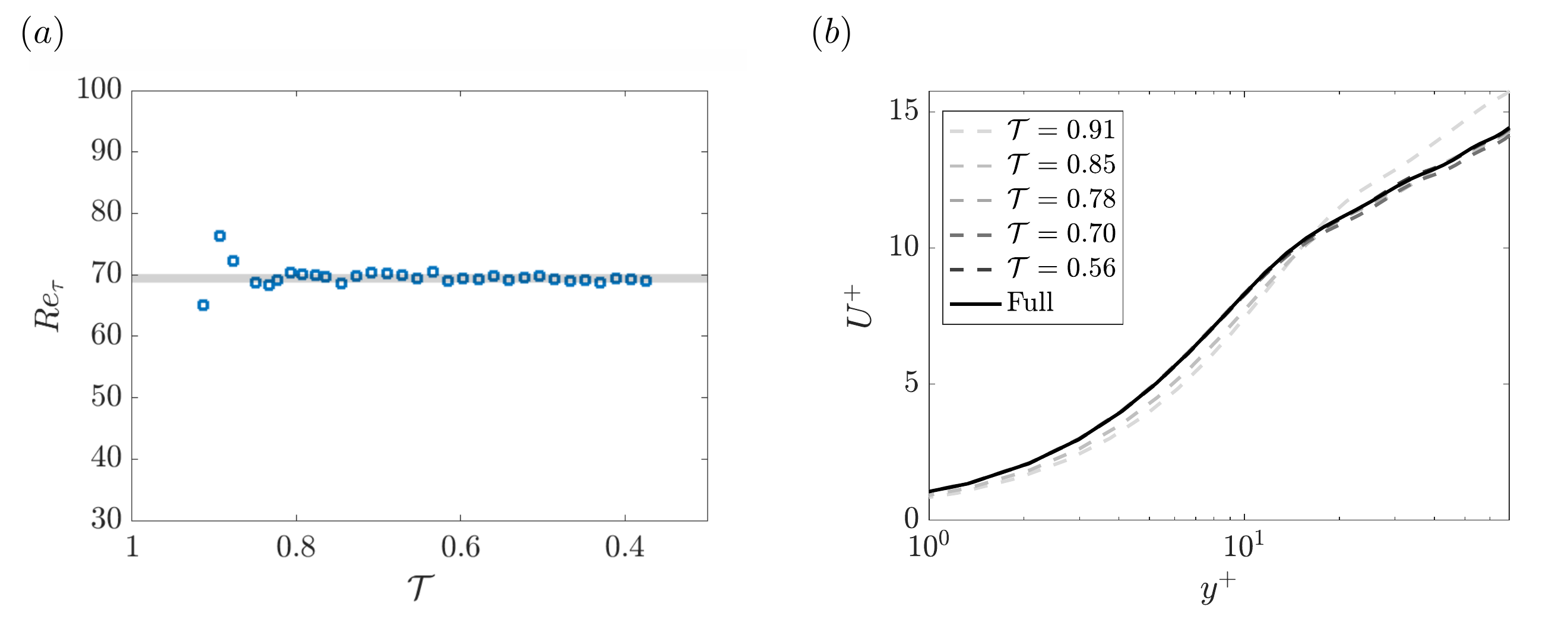}
\caption{Friction Reynolds number (a) and mean flow velocity profiles (B) for GQL models with different controllability threshold. The gray-shaded are in (a) represents the average reference Reynolds number, computed with 95\% confidence level from simulations of the full ROM. Results for five selected controllability thresholds are shown in (b).}
\label{fig:9}
\end{figure}

The convergence of the average friction Reynolds number is found to be significantly faster in the driven GQL approach, as shown in figure \ref{fig:9}(a). The reference $Re_\tau$ is reached for $\mathcal{T} \approx 0.85$, and the values remain virtually constant as $\mathcal{T}$ decreases. In standard GQL, on the other hand, the convergence of $Re_{\tau}$ was found to be more oscillatory, with a slight overestimation up to $\mathcal{T}=0.12$. The same trend is seen in the mean flow profiles, wherein all models with $\mathcal{T} \leqslant 0.85$ present similar levels of agreement with the full ROM. But even when a very severe linearisation is applied, as, for instance, with $\mathcal{T} = 0.91$, the agreement is still reasonably good. This reveals that good estimates of the mean flow can be obtained with a significant degree of model reduction. The sparsity index obtained for $\mathcal{T}=0.91$ is $\mathcal{S}=4.3$.

\begin{figure}
\centering
\includegraphics[trim=10cm 1cm 12cm 1cm, clip=true,width=\linewidth]{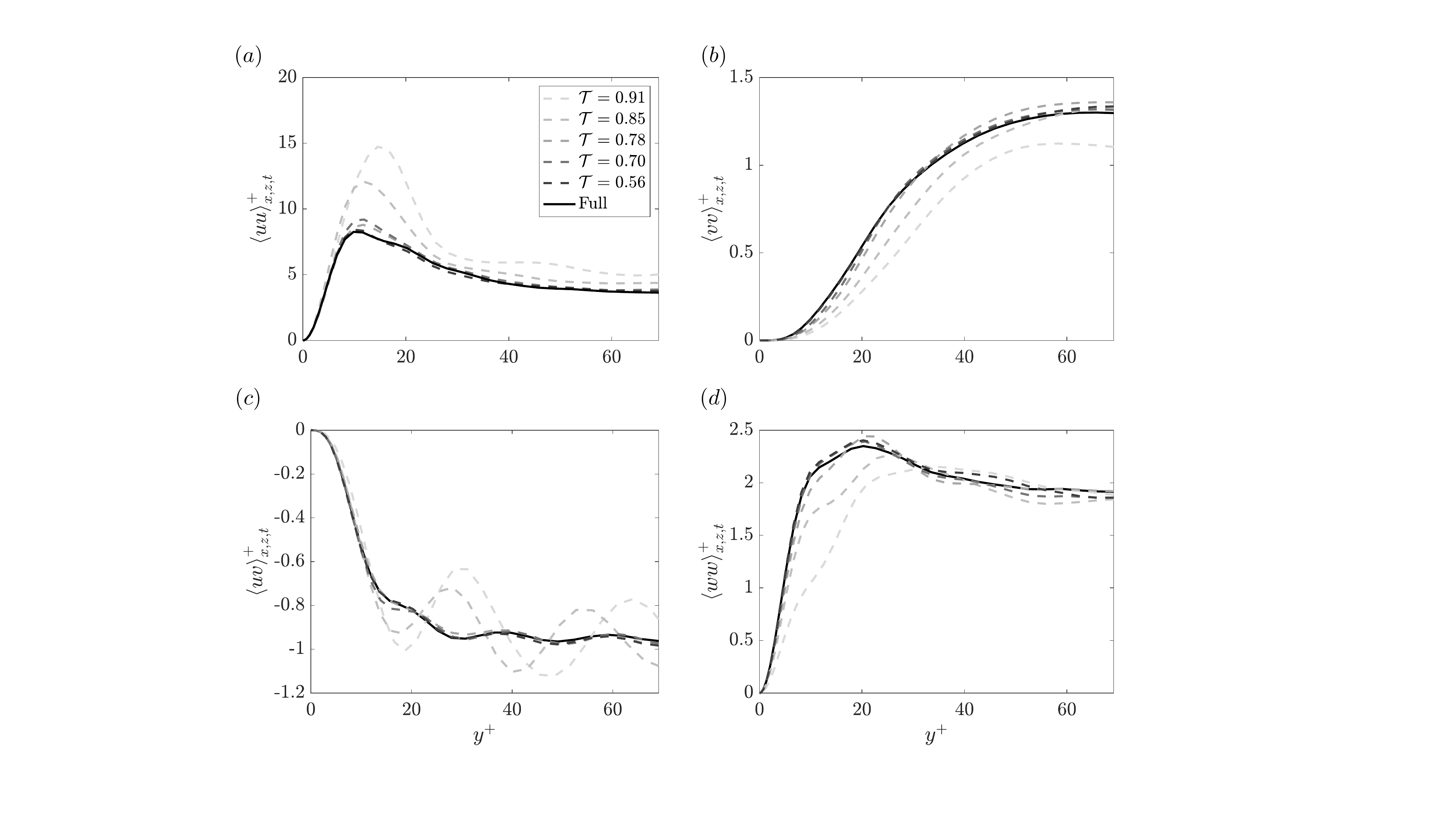}
\caption{Second-order statistics of different driven GQL-ROMs for selected controllability thresholds.}
\label{fig:10}
\end{figure}

\begin{figure}
\centering
\includegraphics[trim=0cm 0cm 0cm 0cm, clip=true,width=0.5\linewidth]{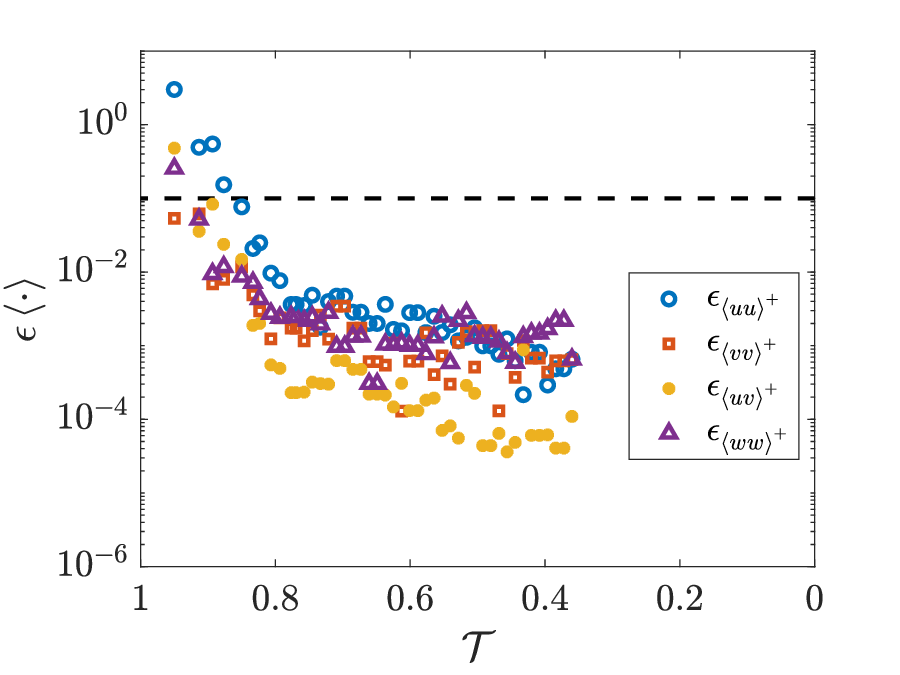}
\caption{Rms errors in second-order statistics of the driven-GQL models with respect to the full ROM.}
\label{fig:11}
\end{figure}

Figure \ref{fig:10} shows second-order statistics of the selected cases. Again we note that convergence is faster, and good levels of agreement are generally obtained at higher controllability thresholds (higher $\mathcal{T}$) with respect to standard GQL. This is confirmed by analysis of the error metric, figure \ref{fig:11}, which reveals that all four errors fall below the accuracy threshold for $\mathcal{T} \approx 0.85$. The sparsity index associated with this model is $\mathcal{S}=2.54$, which is higher than that obtained in the standard GQL framework. This indicates that the driven-GQL framework offers a more accurate description of turbulence dynamics. In appendix \ref{secA1} we present results of driven-GQL ROMs at a higher Reynolds number, $Re=2000$. The trends are similar to those observed at $Re=1000$: convergence in wall-shear stress and second-order statistics is achieved with a similar, large number of linearised modes. We therefore expect significant degrees of linearisation/model reduction degrees of model to also be possible at higher Reynolds numbers.

\section{Discussion}
\label{sec:discussion}

The analysis presented in the previous section revealed that the GQL framework can be associated with ROMs to provide model reduction in systems that are already quite low-dimensional. The approach is similar in spirit to the GQL models explored previously \cite{marston2016generalized,hernandez_yang_hwang_2022,hernandez_yang_hwang_2022_2}, but with a fundamental difference: here linearisation is performed based on the controllability of the modal basis, as opposed to the wavenumber criterion adopted in those studies. This leads to systems in which modes with low and high wavenumbers are linearised together. This is illustrated in table \ref{tab:3}, which presents a description of the number of linearised modes for each wavenumber pair of the basis. Two selected GQL-ROMs are described, derived in the standard and driven frameworks. The linearisation thresholds are chosen such that the two ROMs have similar accuracy, based on the error metric defined previously. In both ROMs linearisation affects all wavenumbers almost uniformly, apart from the mean flow modes ($(k_x, k_z) = (0,0)$), which are less impacted. 

\begin{table}
\centering
\caption{Number of linearised modes for each wavenumber pair, $N_s$, for standard and driven GQL-ROMs with similar levels of accuracy.}
\label{tab:3}       
\begin{tabular}{ccc}
\hline\noalign{\smallskip}
& Driven GQL & Standard GQL \\
\hline\noalign{\smallskip}
$(k_x/\alpha,k_z/\beta)$ & $N_s (\mathcal{T} = 0.85,\mathcal{S}=2.54)$ & $N_s (\mathcal{T} = 0.6,\mathcal{S}=2.01)$\\
\noalign{\smallskip}\hline\noalign{\smallskip}
(0,0) & 6 & 0 \\
(1,0) & 42 & 30 \\
(2,0) & 42 & 30 \\
(0,1) & 42 & 30 \\
(1,-1) & 42 & 30 \\
(2,-1) & 42 & 30 \\
(1,1) & 42 & 30 \\
(2,1) & 42 & 30\\
(0,2) & 42 & 30\\
(1,-2) & 42 & 30 \\
(1,2) & 42 & 30 \\
(2,-2) & 42 & 32\\
(2,2) & 42 & 32 \\
\noalign{\smallskip}\hline
\end{tabular}
\end{table}

\begin{figure}
\centering
\includegraphics[trim=12cm 12cm 11cm 8cm, clip=true,width=\linewidth]{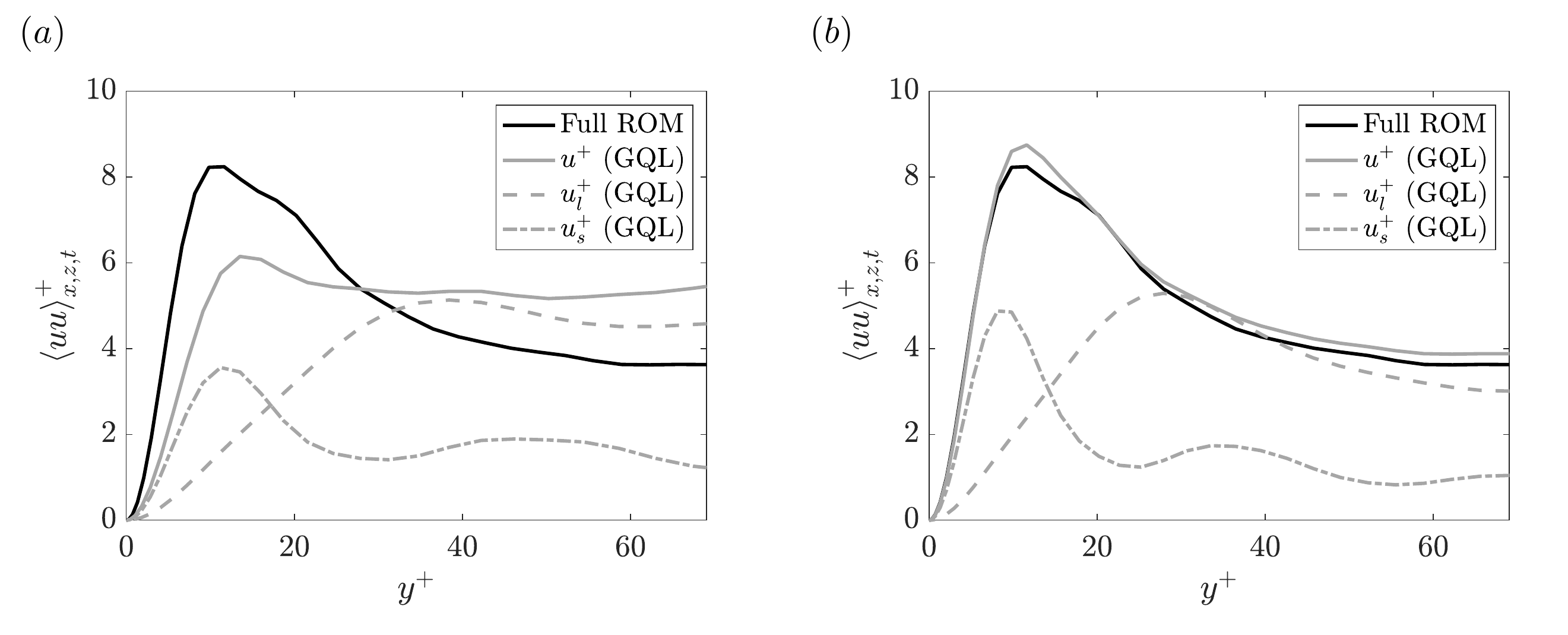}
\caption{$\left<uu\right>^+_{x,z,t}$ component of the Reynolds stress tensor, decomposed into their large, $u^+_l$ and small, $u_s^+$, components. (a) Standard GQL-ROM; (b) Driven GQL-ROM. Linearisation threshold corresponds to $\mathcal{T}=0.76$ in both cases.}
\label{fig:12}
\end{figure}

As discussed above, for this level of accuracy, a larger number of modes is linearised in the driven GQL approach, producing a sparser nonlinear tensor. This suggests that the driving term, $Q_{ijk}a_j a_k \rvert^{s,l,l}$, plays an important role in the dynamics, and its inclusion improves the description of the flow statistics. In order to further investigate this point, we compare how energy is distributed into the large, $u^+_l$, and small, $u^+_s$, components in standard and driven GQL-ROMs derived with the same controllability thresholds. For that we choose $\mathcal{T}=0.76$, a threshold for which the driven GQL-ROM is almost entirely converged, whereas the standard GQL model still has high associated errors. Figure \ref{fig:12} shows profiles of $\left<uu\right>^+_{x,z,t}$, as well as its small and large counterparts. In the standard GQL approach the peak energy of the $u_s^+$ component is diminished with respect to that of the driven GQL model, while that of the $u^+_l$ component is amplified. This indicates that the term $Q_{ijk}a_j a_k \rvert^{s,l,l}$ acts like a source term for $a_j^s$, transferring energy from modes of high controllability to modes of low controllability. Consequently, the net effect of terms $Q_{ijk}a_j a_k \rvert^{l,l,s}$ and $Q_{ijk}a_j a_k \rvert^{l,s,l}$ in the equations for $a_j^l$ is likely that of an energy sink. As these three terms are discarded in the standard GQL approach, the energy of the most controllable modes increases, and $u^+_l$ is overestimated ( in this case being higher than the energy of the full ROM itself). This issue may be directly associated with the overestimation of the $Re_\tau$ observed with the standard GQL model. The generation of wall-shear stress is directly associated with the dynamics of the self-sustaining processes. In small computational boxes/low Reynolds numbers, this occurs essentially in the near-wall region \cite{kravchenko1993relation,choi1994active,orlandi1994generation}. As the Reynolds number is increased, the contribution of self-sustaining processes in the logarithmic region to the wall-shear stress gradually increases \cite{de2016skin}. In the modal basis considered here, the self-sustaining processes involve essentially the large streaks/roll pairs and oblique waves observed in the first controllability modes (see modes in figure \ref{fig:1}(a)-(c)), which are responsible for most of the turbulence production. Their energy indeed peaks further from the wall, as can be seen in figure \ref{fig:12}. And in the standard GQL model, as the energy transfer to less controllable modes is inhibited, the energy of these flow structures is artificially boosted, leading to an overestimation of $Re_\tau$. Moreover, the distribution of $u_l^+$ and $u_s^+$ energies along $y$ has a clearer interpretation in the driven GQL model, with $u_s^+$ more clearly associated with smaller near-wall structures, and $u_l^+$ related to larger structures further from the wall.

\begin{figure}
\centering
\includegraphics[trim=0cm 0cm 0cm 0cm, clip=true,width=0.5\linewidth]{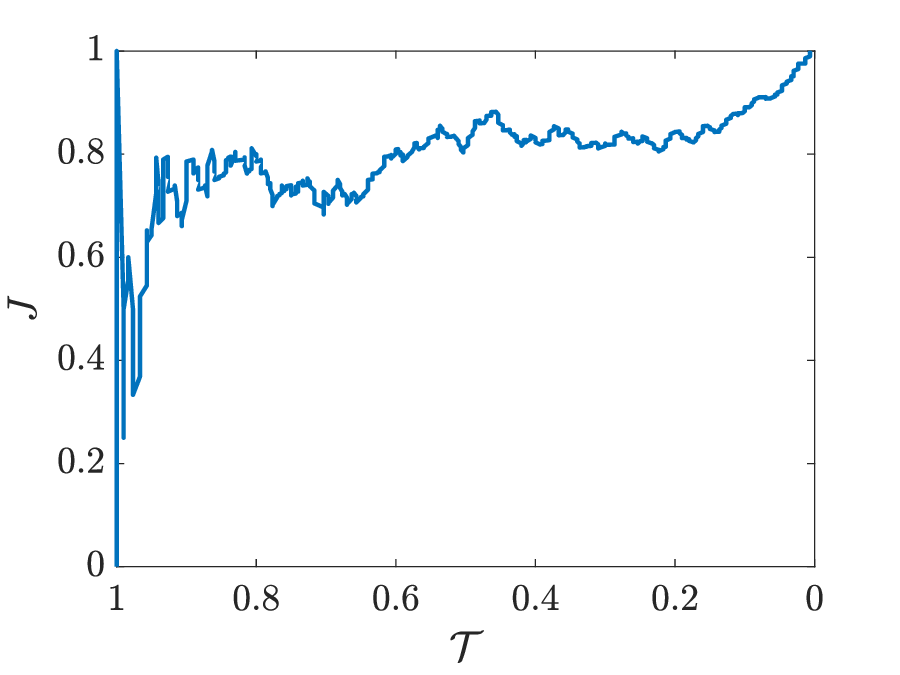}
\caption{Jaccard similarity index between two sets containing the most controllable modes and the most energetic modes, as a function of linearisation threshold.}
\label{fig:13}
\end{figure}

Finally, another point should be made concerning the methodology adopted in the present work. Controllability modes represent the most energetic flow states in a linear system subject to white-noise forcing. This idealised scenario certainly differs, to some extent, from a fully turbulent flow, and one might then question the validity of a controllability criterion to perform linearisation. It is thus instructive to assess the degree to which the most controllable modes (belonging to the $\bu_l$ group and determined \textit{a priori}) correspond to the most energetic modes in the turbulent flow. The latter can be ranked \textit{a posteriori}, according to the average energy of their temporal coefficients, $\overline{a_j^{2}}$. The similarity between the two sets of modes can be characterised by the Jaccard index, which, for two generic sets $X$ and $Y$ takes the form,

\begin{equation}
J(X,Y) = \frac{\left| X \cap Y \right|}{\left| X \cup Y \right|},
\end{equation}
where the symbol $| \cdot|$ denotes the number of elements in the union and intersection of the sets. By definition, $0 \leqslant J \leqslant 1$, and $J(X,Y)=0$ if the intersection between $X$ and $Y$ is empty. Figure \ref{fig:13} shows the Jaccard index between the most controllable and the most energetic modes for different linearisation thresholds explored in the present study. The most energetic modes are determined from a simulation of the full ROM. Naturally, as $\mathcal{T}$ decreases, less modes are linearised, and the two sets become increasingly identical. Notice, however, that $J$ increases sharply from $\mathcal{T}=1$ to $\mathcal{T}=0.85$. With $\mathcal{T}=0.85$ there is already 80\% similarity between the two sets, despite them having relatively few modes (at this threshold, there are 90 modes in the $\bu_l$ set and 510 in the $\bu_s$ set, see table \ref{tab:3}). This sheds some light on the success of driven GQL-ROMs in providing accurate estimates of the flow statistics with relatively severe linearisation thresholds. If we consider the structure of the neglected terms, $Q_{ijk}a_j a_k \rvert^{s,s,s}$, it follows that they possess small amplitudes if the modes involved have low energy, due to the product $a_ja_k$. Therefore, discarding them will have limited impact on the dynamics.

\section{Conclusion}
\label{sec:conclusion}

In this work, we have applied generalised quasilinear approximations to turbulent plane Couette flow. The GQL framework is coupled with a recently-developed ROM \cite{Cavalieri&Nogueira_PRF2022} that uses controllability modes of the linearised Navier-Stokes system as basis functions. The methodology is conceptually similar to that used previously in rotating Couette flow \cite{tobias2017three} and turbulent channel flow \cite{hernandez_yang_hwang_2022,hernandez_yang_hwang_2022_2}, but with a different choice of linearisation criterion. Instead of decomposing the velocity field into low- and high-wavenumber Fourier components, here we split the basis into groups of high and low controllability. The equations of the latter are then linearised. Furthermore, high-controllability modes are seen to correspond, to a great extent, to the most energetic turbulent flow structures. The framework explored here allows linearisation to be done \textit{a priori}, without the need to run calculations of the full system. Furthermore, another novelty of the present work is a new GQL framework, referred to as driven-GQL, wherein the equations for the low-controllability modes are driven by a term stemming from nonlinear interactions of the high-controllability group; in standard GQL approximations, this \enquote{driving} term is discarded. GQL-ROMs successfully recover relevant statistics of the full model with a relatively high controllability threshold, leading to a large number of linearised modes. Driven GQL models were found to converge more rapidly to the full system than standard GQL approximations, due to the energy transfer from high-controllability modes to low-controllability modes which is taken into account in this approach. This mechanism was found to be important for an accurate calculation of the wall shear-stress and the Reynolds stress tensor, and therefore should be considered in future GQL models.

Overall, the results of the present study show that it is possible to obtain further model reductions in ROMs that are already quite compact. In both GQL approaches explored, we were able to discard more than half of the nonlinear interactions without significantly altering the flow statistics. Some minimal amount of nonlinearity must be kept, though, as models with highly-truncated nonlinear terms were seen to yield large errors. This work also suggests new directions for the study of dominant non-linear interactions among coherent structures in turbulence, as the GQL reduced-order models isolate the key interactions that lead to observed statistics. It is worth mentioning that the GQL framework laid out here is not restricted to a basis of controllability modes. It can be explored with any choice of orthogonal basis functions (such as POD modes, balanced modes, Stokes modes, etc.), provided a suitable linearisation criterion is defined.

The methodology laid out here can be used in future studies to extend ROMs to larger Reynolds numbers. As the Reynolds number is increased, a larger basis is required for an accurate representation of turbulence statistics. The most computationally-expensive operations in such models involve the nonlinear tensor, which has dimensions of $N^3$. But as seen here, the GQL framework leads to sparser nonlinear tensors, which can potentially provide significant reductions in computational cost at large Reynolds number. It is worth pointing out that, for parallel flows, as is the case here, the the quadratic term in standard (non-GQL) ROMs is already quite sparse. Since we work in the Fourier space, many of the coefficients of the nonlinear term are identically zero due to wavenumber triad rules. However, this is not the case for ROMs of flows with inhomogeneous directions. An example is the ROM for a 2D mixing layer recently developed by \citet{cavalieri_aiaa2023}, for which the quadratic coefficients form a full tensor. In that case, the GQL-ROM framework offers a much higher potential for sparsification of the nonlinear term, and we therefore hope that the methodology developed here will allow even more significant model reductions for such flows.

\backmatter

\bmhead{Acknowledgments}

The authors would like to thank prof. Greg Chini for helpful discussions on generalised quasilinear approximations.

\bmhead{Funding}
This work was funded by the S\~ao Paulo Research Foundation-FAPESP through grant No 2022/06824-4 and by Conselho Nacional de Desenvolvimento Tecnol\'ogico-CNPq through grant No.313225/2020-6.

\bmhead{Statements and Declarations}
The authors report no conflict of interest.

\begin{appendices}
\section{Driven-GQL ROMs at higher $Re$}\label{secA1}

\begin{figure}
\centering
\includegraphics[trim=12cm 12cm 11cm 8cm, clip=true,width=\linewidth]{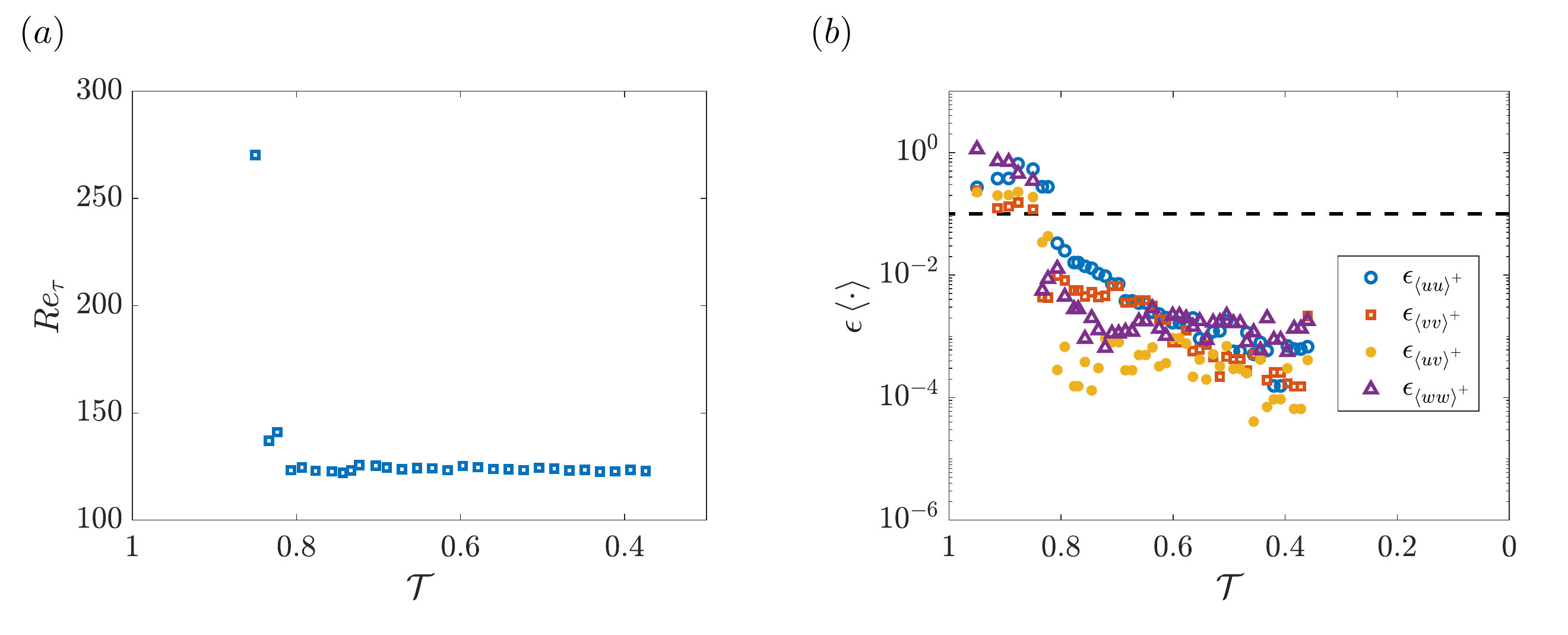}
\caption{Driven-GQL ROMs at $Re=2000$ for different linearisation thresholds. (a) Friction Reynolds number; (b) RMS errors in second-order statistics.}
\label{fig:14}
\end{figure}

In this section we present results of driven-GQL models simulated for $Re=2000$, using the same $N=600$ ROM. Figure \ref{fig:14} shows the friction Reynolds number and rms errors in second-order statistics for different linearisation thresholds. Notice that the $Re_{\tau}$ and the second-order statistics of the GQL-ROMs converge to that of the full ROM with a degree of linearisation similar to that observed at $Re=1000$, indicating that significant model reductions can also be expected at higher Reynolds numbers.

\end{appendices}

\bibliography{bibfile}

\end{document}